\documentclass[3p,times]{elsarticle}

\usepackage{ecrc}

\volume{00}

\firstpage{1}

\journalname{}

\runauth{Jun-Sik Sin et al}

\jid{}

\jnltitlelogo{}

\usepackage{amssymb}
\usepackage[figuresright]{rotating}

\large
\begin{document}

\begin{frontmatter}



\dochead{}
%% e.g. \dochead{17th International Conference on Dynamical Processes in Excited States of Solids}

\title{Effect of solvent polarization on electric double layer of a charged soft surface in an electrolyte solution}
\author{\large Jun-Sik Sin}
\large
\ead{ryongnam24@yahoo.com}
\author{\large Nam-Hyok Kim}
\author{\large Chung-Sik Sin}
\address{\large Department of Physics, \textbf{Kim Il Sung} University, Taesong District, Pyongyang, DPR Korea}

\linespread{1.6}
\begin{abstract}
\large

We study the electric double layer near a charged soft surface by using a mean-field approach including non-uniform size effect and solvent  polarization. Based on a free energy model, electrostatic potential and number densities of water and ions are obtained numerically by solving coupled differential equations including the Poisson’s equation. 
We find that the solvent polarization significantly affects electrostatic potential, ion number densities and permittivity in an electrolyte solution near a charged soft surface. We prove that the consideration of the solvent polarization increases the magnitude of Donnan potential. We also demonstrate that within the fixed charge layer the number densities of counterions and coions are slightly smaller than when the solvent polarization is not considered.  
An increase in the ionic size enhances the influence of solvent polarization on the electrostatic properties within the fixed charged layer of charged soft surface. In the region of high surface-charge-density, the permittivity and the number density of water dipoles considerably decrease due to water polarization. 
\end{abstract}

\large
\end{frontmatter}

% \linenumbers

\section{Introduction}
\large
Electric double layer electrostatics of charged soft surface in an electrolyte solution is necessary for studying important biological and chemical applications such as the electrokinetics of biological cells \cite%
{Ohshima_STAM_2011}, the effect of electric double layer in bacterial adhesion to surfaces \cite%
{Bos_SSR_2002, Kerchove_Langmuir_2005}, charging and swelling of cellulose films \cite%
{Werner_JCIS_2007}, injectable material for local delivery of therapeutic agents \cite%
{Xing_AdvMater_2016} and highly efficient dye absorbents for wastewater treatment \cite%
{Zhao_SciRep_2017}, etc. 

H. Ohshima initially developed a Poisson-Boltzmann approach to describe electric double layer electrostatics of charged soft surface with a structural modeling, where a  charged soft surface consists of a surface charge layer with a uniform density N of ionized groups of valence Z (the fixed charge layer of FCL) and a uncharged hard plate \cite%
{Ohshima_JCIS_2003, Ohshima_Book_2006, Ohshima_JCIS_2008_1,  Ohshima_JCIS_2008_2,Ohshima_STAM_2009, Ohshima_SoftMatter_2012}. This model for charged soft surface were used to study several issues such as electrostatic interactions between two similar or dissimilarly soft charged interfaces \cite%
{Ohshima_CSA_2011} and electrophoretic mobility of soft particles \cite%
{Ohshima_JPCA_2012}, and other related electrokinetic phenomena \cite%
{Ohshima_COCIS_2012}
, etc.
 
Further modifications of mean-field approaches for a charged soft surface  were developed by accounting for finite ion size \cite%
{Das_PRE_2014, Das_RA_2015} and pH dependent
charge density\cite%
{ Das_RscAdv_2015, Das_JAP_2015}.

In fact, for the case of a hard charged surface in an electrolyte,  many researchers used the Possion-Boltzmann approach for supporting peculiar phenomena of electrochemistry, colloid science and biophysics for more than a century\cite%
{Gouy_JPhysF_1910, Chapman_PhilosMag_1913}. Some alterations were made from taking into account finite ion size \cite%
{Bikerman_PhilosMag_1942, Wicke_ZEC_1952, Iglic_JPhysF_1996, Andelman_PRL_1997, Andelman_EA_2000, Iglic_EA_2001, Iglic_Bioelechem_2005}. In particular, during the past decade, mean-field approaches accounting for non-uniform size effect of ions \cite%
{Chu_Biophys_2007, Kornyshev_JPCB_2007, Biesheuvel_JCIS_2007, Li_PRE_2011,  Li_PRE_2012, Boschitsch_JCC_2012, Siber_PRE_2013}   were developed to explain the experimental facts such as asymmetry of properties of electric double layer and counterion stratification in multicomponent electrolyte solution. 

In order to present a reasonable permittivity formula of electrolyte solution near a charged surface, the authors of \cite%
{Iglic_Bioelechem_2010, Iglic_Bioelechem_2012} developed a mean-field approach taking into account excluded volume effect and water polarization. The approach is based on the assumption that each ion and water molecule have the same size. 

In order to overcome the excessive
 assumption and describe asymmetric properties of electric double layer around a charged surface, we proposed a mean-field approach taking into account both non-uniform size effect and water polarization with assumption of small ion volume \cite%
{Sin_EA_2015, Sin_PCCP_2016}.  

Using a lattice statistics, Gongadze and Iglic \cite%
{Iglic_EA_2015} presented an analytical mean-field approach describing electric double layer near a charged surface in contact with a binary electrolyte solution without the assumption of small ion volume used in \cite%
{Sin_EA_2015}. But the approach is not extended to the study of a multicomponent electrolyte solution because the generalization of the approach cannot predict the peculiar phenomena occurring in multicomponent electrolyte solution such as counterion stratification  \cite%
{Biesheuvel_JCIS_2007, Li_PRE_2011, Li_PRE_2012, Boschitsch_JCC_2012} and asymmetric depletion of cosolvents \cite%
{Andelman_JPCB_2009}. 

Very recently, we updated the approach of \cite%
{Sin_EA_2015} with a different entropy representation \cite%
{Sin_EA_2016}.  Providing simple equations compared to that of the previous approach \cite%
{Sin_EA_2015}, the approach naturally predict the phenomena such as counterion stratification  \cite%
{Sin_EA_2016} and asymmetric solvent depletion \cite%
{Sin_PCCP_2016_2} near a charged surface in a multicomponent electrolyte solution. Although solvent polarization and non-uniform size effect were considered for the case of a charged hard surface, the issue remains unsolved for a charged soft surface. 

	 In this paper, we present a mean-field theory to study the effect of solvent polarization in the electrostatics of a charged soft surface immersed in an electrolyte solution. Our theory is based on a free energy model accounting for orientational ordering of water dipoles as well as the difference in size between ions and water molecules and the geometry of the charged soft surface (see Fig. 1). 
	The main result is that for the case when considering orientational ordering of water dipoles, the Donnan potential is higher than when orientational ordering of water dipoles is not considered, whereas the number densities of both counterions and coions get slightly small. We indicate nature of such an increase in the Donnan potential when considering orientational ordering of water dipoles. We also show that the permittivity and number density of water dipoles are substantially affected by the steric factor and surface-charge-density. In practice, our approach and results will be applied to many applications using a biomembrane being accepted as the ion-selective interface of the FCL and the adjacent electrolyte \cite%
{Ohshima_STAM_2011, Ohshima_Biophys_1985, McLaughlin_ARBBC_1989}
. 

\section{Theory}
\begin{figure}
\begin{center}
\includegraphics[width=0.6\textwidth]{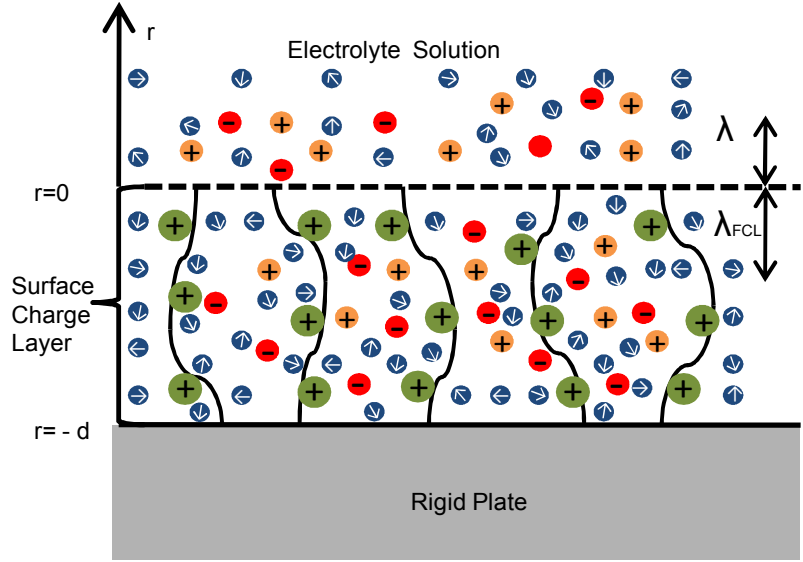}
\caption{\large (Color online) Schematic illustration of the charged soft surface(consisting of a fixed charge layer (FCL) and a rigid plate) in contact with an electrolyte solution. The FCL contains water dipoles(shown in blue), cations(shown in yellow) and anions(shown in red)  as well as another kind of ions (shown in green) located within the FCL with a valence Z  and a uniform charge density N.}
\label{fig:1}
\end{center}
\end{figure}

\large
We consider an electrolyte solution consisting of ions and water molecules in contact with a charged soft surface. As in the previous study \cite%
{Das_PRE_2014}, the charged soft surface is replaced by a hard plate covered by a layer of ion-penetrable polyelectrolytes, shown in Fig. \ref{fig:1}.  

The free energy of the total system  $F$ is given as a functional of the local electrostatic potential   $\psi\left(r\right)$ and the number densities of different ionic species $n_{+}$, $n_{-}$  and the number density of water molecules 
$n_{w}\left(r\right)=\left<\rho\left(\omega, r \right)\right>_{\omega}$
 \begin{eqnarray}
	F=\int{d{\bf r}}\left(-\frac{\varepsilon_{0}\varepsilon E^2}{2}+e_{0}z\psi\left(n_{+}-n_{-}\right)+\left<\rho\left(\omega\right)\gamma{p_{0}}E\cos\left(\omega\right)\right>_{\omega}+e_{0}ZN\psi\left[\theta\left(r+d\right)-\theta\left(r\right)\right]\right)\nonumber\\+ \int{d{\bf r}}\left(-\mu_{+}n_{+}-\mu_{-}n_{-}-\left<\mu_{w}\left(\omega\right)\rho\left(\omega\right)\right>_{\omega}-Ts\right),
\label{eq:1}
\end{eqnarray}
where $\left<f\left(\omega\right)\right>_{\omega}={\int f\left(\omega\right)2\pi\sin\left(\omega\right)  d\omega}/\left(4 \pi\right)$ in which $\omega$ represents the angle between the vector {\bf p} and the normal to the charged surface.  Here {\bf p} and {\bf E} are the dipole moment of water molecules and the electric field strength, respectively. $z$ is the ionic valence of ions in the electrolyte solution.

The first term means the self energy of the electrostatic field, where $\varepsilon_{0}$ indicates vacuum permittivity. $\varepsilon$ is equal to $n^2$, where $n=1.33$ is the refractive index of water. The second and third terms represent the electrostatic energy of the ions and water dipoles in the electrolyte solution, respectively.  Here  $\gamma=\left(2+n^2\right)/2$,  $p_{0}=\left|{\bf p}\right|$  and $E=\left|{\bf E}\right|$ and $e_{0}$ is the elementary charge. The fourth term represents the electrostatic energy of the ions fixed within the FCL and the function $\theta$ is the Heaviside function.  
The next terms couple the system to a bulk reservoir, where $\mu_{+}$ and $\mu_{-}$ mean the chemical potentials of positive and negative ions and $\mu_{w}\left(\omega\right)$ corresponds to the chemical potential of water dipoles with orientational angle $\omega$.
$T$ and $s$ are the temperature and the entropy density, respectively.

 Due to the same reason as in \cite%
{Das_RA_2015}, excluded volume effect of polyelectrolytes in FCL can be neglected. 
As a consequence, we get the Lagrangian of the electrolyte using the method of undetermined multipliers as in \cite%
{Sin_EA_2016}, 
 \begin{eqnarray}
	L=F-\int\lambda\left({\bf r}\right)\left(1-n_{+}V_{+}-n_{-}V_{-}-n_{w}V_{w}\right)d{\bf r},
\label{eq:6}
\end{eqnarray}
where $\lambda$ is a local Lagrange parameter and the negative and positive ion and water molecule occupy volumes of $V_{+}$, $V_{-}$ and $V_{w}$, respectively.

When the reference point of the electric potential is located at $r=+\infty$, $\psi\left(r\rightarrow+\infty\right)=0$.  Other boundary conditions are  $n_{+}\left(r\rightarrow+\infty\right)=n_{+b}$,  $n_{-}\left(r\rightarrow+\infty\right)=n_{-b}$,  $n_{w}\left(r\rightarrow+\infty\right)=n_{wb}$ and $\lambda\left(r\rightarrow+\infty\right)=\lambda_{b}$, where  $n_{+b}$, $n_{-b}$, $n_{wb}$ and $\lambda_{b}$  represent the number densities of positive and negative ions and water molecules, and the Lagrange parameter at  $r\rightarrow+\infty$, respectively. 

Considering the boundary conditions,  we get the below equations by using the same method as in \cite%
{Sin_EA_2016}
\begin{equation}
	n_{+}=\frac{n_{+b}\exp\left(-V_{+}h-e_{0}z\beta\psi\right)}{n_{+b}V_{+}\exp\left(-V_{+}h-e_{0}z\beta\psi\right)+n_{-b}V_{-}\exp\left(-V_{-}h+e_{0}z\beta\psi\right)+n_{wb}V_{w}\exp\left(-V_{w}h\right)\left\langle\exp\left(-\gamma p_{0}\beta E\cos\left(\omega\right)\right)\right\rangle_{\omega}},   
\label{eq:14a}
\end{equation}
\begin{equation}
	n_{-}=\frac{n_{-b}\exp\left(-V_{-}h+e_{0}z\beta\psi\right)}{n_{+b}V_{+}\exp\left(-V_{+}h-e_{0}z\beta\psi\right)+n_{-b}V_{-}\exp\left(-V_{-}h+e_{0}z\beta\psi\right)+n_{wb}V_{w}\exp\left(-V_{w}h\right)\left\langle\exp\left(-\gamma p_{0}\beta E\cos\left(\omega\right)\right)\right\rangle_{\omega}},  
\label{eq:14b}
\end{equation}
\begin{equation}
	n_{w}=\frac{n_{wb}\exp\left(-V_{w}h\right)\langle\exp\left(-\gamma p_{0}\beta E\cos\left(\omega\right)\right)\rangle_{\omega}}{n_{+b}V_{+}\exp\left(-V_{+}h-e_{0}z\beta\psi\right)+n_{-b}V_{-}\exp\left(-V_{-}h+e_{0}z\beta\psi\right)+n_{wb}V_{w}\exp\left(-V_{w}h\right)\left\langle\exp\left(-\gamma p_{0}\beta E\cos\left(\omega\right)\right)\right\rangle_{\omega}},
\label{eq:15}
\end{equation}
\begin{eqnarray}
n_{+b}\left(\exp\left(-V_{+}h-e_{0}z\beta\psi\right)-1\right)+n_{-b}\left(\exp\left(-V_{-}h+e_{0}z\beta\psi\right)-1\right)+n_{wb}\left(\exp\left(-V_{w}h\right)\frac{\sinh\left(\gamma p_{0}\beta E\right)}{\gamma p_{0}\beta E}-1\right)=0.
\label{eq:16}
\end{eqnarray}
where $N_{b}=n_{+b}+n_{-b}+n_{wb}$,  $h\equiv\lambda-\lambda_{b}$ and
\begin{eqnarray}
	\left\langle\exp\left(-\gamma p_{0}\beta E\cos\left(\omega\right)\right)\right\rangle_{\omega}=\frac{2 \pi\int^{0}_{\pi}d\left(cos\left(\omega\right)\right)\exp\left(-\gamma p_{0}\beta E\cos\left(\omega\right)\right)}{4\pi}=\frac{\sinh\left(\gamma p_{0} E \beta\right)}{\gamma p_{0} E \beta}.
\label{eq:11}
\end{eqnarray}

The Euler$-$Lagrange equation for  $\psi\left(r\right)$  yields the Poisson equation
\begin{eqnarray}
	\nabla\left(\varepsilon_{0}\varepsilon_{r}\nabla\psi\right)=-e_{0}z\left(n_{+}-n_{-}\right),    r>0 \nonumber\\
	\nabla\left(\varepsilon_{0}\varepsilon_{r}\nabla\psi\right)=-\left[e_{0}z\left(n_{+}-n_{-}\right)+e_{0}ZN\right],   -d< r<0	
\label{eq:19}
\end{eqnarray}
where
 \begin{eqnarray}
	\varepsilon_{r} \equiv n^2+\frac{n_{w}\left(r\right)\left(\frac{2+n^2}{3}\right)p_{0}\mathcal{L}\left(\gamma{p_{0}}E\beta \right)}{\varepsilon_{0}E}.
\label{eq:20}
\end{eqnarray}
where a function $\mathcal{L}\left(u\right)=\coth\left(u\right)-1/u$ is the Langevin function, $k_{B}$  is the Boltzmann constant and $\beta=1/\left(k_{B}T\right)$.

We make the assumption that coions and counterions have the same size, where  $V_{+}=V_{-}, V_{+}\neq V_{w}$. This assumption is based on the fact that the properties of electric double layer is hardly affected by coion size \cite%
{Sin_EA_2015}. On the other hand, for the neutrality of electricity in an electrolyte solution, $n_{+b}=n_{-b}=n_{b}$.
Using $n_{+b}=n_{-b}=n_{b}$ and $V_{+}=V_{-}=V$, Eqs. (\ref{eq:14a}), (\ref{eq:14b}), (\ref{eq:15}), (\ref{eq:16})  are rewritten as follows:
\begin{eqnarray}
		n_{+}=\frac{n_{b}\exp\left(-e_{0}z\beta\psi\right)}{ 2n_{b}V\cosh\left(e_{0}z\beta\psi\right)+n_{wb}V_{w}\left\langle\exp\left(-\gamma p_{0}\beta E\cos\left(\omega\right)\right)\right\rangle_{\omega}\exp\left(\left(V-V_{w}\right)h\right)},  
\label{eq:22a}
\end{eqnarray}
\begin{eqnarray}
		n_{-}=\frac{n_{b}\exp\left(+e_{0}z\beta\psi\right)}{ 2n_{b}V\cosh\left(e_{0}z\beta\psi\right)+n_{wb}V_{w}\left\langle\exp\left(-\gamma p_{0}\beta E\cos\left(\omega\right)\right)\right\rangle_{\omega}\exp\left(\left(V-V_{w}\right)h\right)},  
\label{eq:22b}
\end{eqnarray}
\begin{eqnarray}
		n_{w}=\frac{n_{wb}\left\langle\exp\left(-\gamma p_{0}\beta E\cos\left(\omega\right)\right)\right\rangle_{\omega}\exp\left(\left(V-V_{w}\right)h\right)}{ 2n_{b}V\cosh\left(e_{0}z\beta\psi\right)+n_{wb}V_{w}\left\langle\exp\left(-\gamma p_{0}\beta E\cos\left(\omega\right)\right)\right\rangle_{\omega}\exp\left(\left(V-V_{w}\right)h\right)},  
\label{eq:23}
\end{eqnarray}
\begin{eqnarray}
		\frac{dh}{dr}=\frac{2n_{b}\sinh\left(e_{0}\beta z\psi\right)e_{0}\beta z\frac{d\psi}{dr}+n_{wb}\exp\left(\left(V-V_{w}\right)h\right)\mathcal{L}\left(\gamma p_{0}\beta E\right)\gamma p_{0}\beta\frac{dE}{dr}}{ 2n_{b}V\cosh\left(e_{0}z\beta\psi\right)+n_{wb}V_{w}\left\langle\exp\left(-\gamma p_{0}\beta E\cos\left(\omega\right)\right)\right\rangle_{\omega}\exp\left(\left(V-V_{w}\right)h\right)}.  
\label{eq:24}
\end{eqnarray}

Using Eqs. (\ref{eq:22a}), (\ref{eq:22b}) in Eqs. (\ref{eq:19}) and (\ref{eq:24}), we finally get the equations determining the electrostatic potential as
\begin{eqnarray}
\frac{d}{d\overline{r}}\left(\overline{\varepsilon_{r}}\frac{d\overline{\psi}}{d\overline{r}}\right)=\eta\frac{\sinh\left(\overline{\psi}\right)}{\frac{\sinh\left(\chi\overline{E}\right)}{\left(\chi\overline{E}\right)}\exp\left(\left(1-\frac{1}{\overline{V}}\right)\overline{h}\right)+2\overline{n}_{b}\overline{V}\cosh\left(\overline{\psi}\right)},    \overline{r}>0  \nonumber\\
	\frac{d}{d\overline{r}}\left(\overline{\varepsilon_{r}}\frac{d\overline{\psi}}{d\overline{r}}\right)=\eta\frac{\sinh\left(\overline{\psi}\right)}{\frac{\sinh\left(\chi\overline{E}\right)}{\left(\chi\overline{E}\right)}\exp\left(\left(1-\frac{1}{\overline{V}}\right)\overline{h}\right)+2\overline{n}_{b}\overline{V}\cosh\left(\overline{\psi}\right)}-\frac{1}{K^{2}_{\lambda}},    0>\overline{r}>-\overline{d}
\label{eq:25}
\end{eqnarray}
\begin{eqnarray}
	\frac{d\overline{h}}{d\overline{r}}=\frac{2\overline{n}_{b}\sinh\left(\overline{\psi}\right)\frac{d\overline{\psi}}{d\overline{r}}+\exp\left(\left(1-\frac{1}{\overline{V}}\right)\overline{h}\right)\mathcal{L}\left(\chi \overline{E}\right)\chi\frac{d\overline{E}}{d\overline{r}}}{\frac{\sinh\left(\chi\overline{E}\right)}{\left(\chi\overline{E}\right)}\exp\left(\left(1-\frac{1}{\overline{V}}\right)\overline{h}\right)+2\overline{n}_{b}\overline{V}\cosh\left(\overline{\psi}\right)}-\frac{1}{K^{2}_{\lambda}},
\label{eq:26}
\end{eqnarray}
,where $\overline{r}=r/\lambda$, $\overline{\psi}=e_{0}z\beta \psi$, $\overline{d}=d/\lambda$, $\overline{\varepsilon_{r}}=\varepsilon_{r}/\varepsilon_{p}$,  $\lambda=\sqrt{\frac{\varepsilon_{0}\varepsilon_{p}k_{B}T}{2n_{b}e^{2}_{0}z^{2}}}$, $k_{\lambda}=\frac{\lambda_{FCL}}{ \lambda}$, $\lambda_{FCL}=\sqrt{\frac{\varepsilon_{0}\varepsilon_{p}k_{B}T}{Ne^{2}_{0}zZ}}$, $\overline{h}=hV$, $\overline{V}=V/V_{w}$, $\eta=\frac{1}{n_{wb}V_{w}}$,
$\overline{n}_{b}=\frac{n_{b}}{n_{wb}}$, $\gamma p_{0}\beta E=\gamma p_{0} \sqrt{\frac{2n_{b}}{\varepsilon_{0}\varepsilon_{p}k_{B}T}}\frac{d\overline{\psi}}{d\overline{r}}=\chi\frac{d\overline{\psi}}{d\overline{r}}=\chi\overline{E}$ and $\varepsilon_{p}$ means the relative permittivity of the bulk electrolyte solution.
For the sake of simplicity, we have considered identical values of the steric factor for outside and inside the FCL, although we can easily amend the free energy expression to account for different values of steric factors. We numerically solve Eqs. (\ref{eq:25}), (\ref{eq:26}) with the help of the following dimensionless boundary conditions:
\begin{eqnarray}
	\left(\frac{d\overline{\psi}}{d\overline{r}}\right)_{\overline{r}=-\overline{d}}=0,  \overline{\psi}_{\overline{r}=0^{+}}=\overline{\psi}_{\overline{r}=0^{-}}, \nonumber\\
	\left(\frac{d\overline{\psi}}{d\overline{r}}\right)_{\overline{r}=0^{+}}=\left(\frac{d\overline{\psi}}{d\overline{r}}\right)_{\overline{r}=0^{-}}, \left(\frac{d\overline{\psi}}{d\overline{r}}\right)_{\overline{r}-\rightarrow \infty}=0,
\label{eq:27}
\end{eqnarray}

For the case when considering both finite size effects and orientational ordering of water dipoles, we express the charge neutrality at $\overline{r}=-\overline{d}$ so that 
(considering the case where $d>>\lambda$, $\overline{\psi}_{\overline{r}=-\overline{d}}=\overline{\psi}_{D}$, $\overline{E}_{\overline{r}=-\overline{d}}=0$)
\begin{eqnarray}
	\left[e_{0}NZ+e_{0}z\left(n_{+}-n_{-}\right)\right]_{r=-d}=0,	
	\label{eq:28}
\end{eqnarray}
\begin{eqnarray}
	e_{0}{N}Z=\frac{2e_{0}z n_{b}\sinh\left(\overline{\psi}_D\right)}{\left(1-2n_{b}V\right)\exp\left(\left(1-\frac{1}{\overline{V}}\right)\overline{h}\right)+2n_{b}V\cosh\left(\overline{\psi}_D\right)}.
\label{eq:29}
\end{eqnarray}
In order to determine the electrostatic potential and the number densities of the ions and water molecules, the coupled differential equations including Eqs. (\ref{eq:25}), (\ref{eq:26}) and (\ref{eq:27}) should be numerically solved.

\section{Results and Discussion}
\large

Using the fourth order Runge-Kutta method, we obtain  $n_{+}$, $n_{+}$, $n_w$, $\overline{\psi}$ by combining  Eqs. (\ref{eq:25}), (\ref{eq:26}), (\ref{eq:27}) with boundary conditions. In all the calculations, the temperature T and the number density of water molecules in the bulk electrolyte are taken as $298.15K$ and  $55mol/l$, respectively.
As in \cite%
{Iglic_Bioelechem_2012, Sin_EA_2015, Iglic_EA_2015}, the water dipole moment should be 3.1D so that far away from the charged surface,  $\overline{\psi}\left(r\rightarrow\infty\right)=0$  and the relative permittivity of the electrolyte is equal to the value of pure water, 78.5.    Although the dipole moment of a water molecule in the calculation is higher than the dipole moment of an average water molecule in bulk solution(2.6D) but the value can represent the permittivity of bulk solution(78.5) accounting for cavity field within a mean-field approach using lattice statistics.

\begin{figure}
\includegraphics[width=1\textwidth]{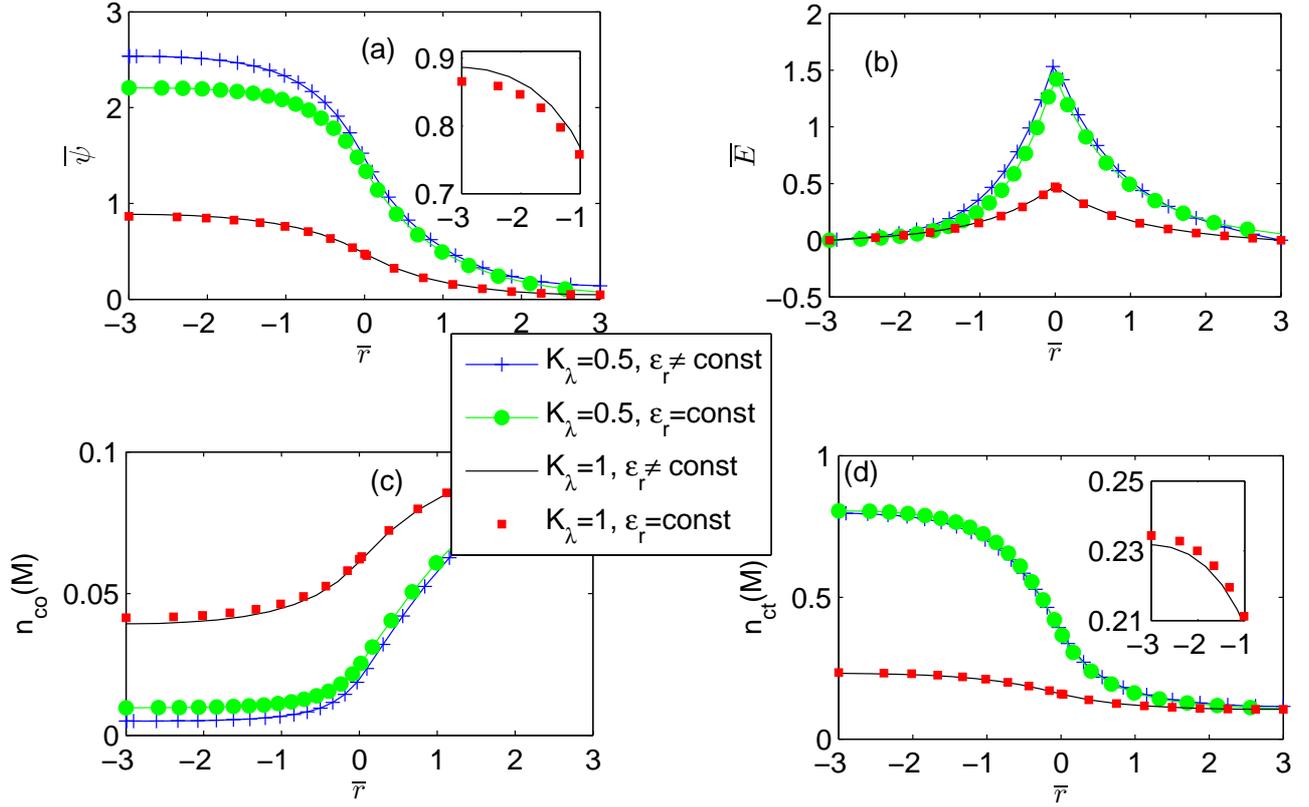}
\caption{\large (Color online) The dimensionless electric potential (a),  dimensionless electric field strength (b), the number density of coions (c) and the number density of counterions (d) near the charged soft surface, when considering or not orientational ordering of water dipoles. The dimensionless  thickness of ion-penetrable polyelectrolytes, the volume of each ion and ionic concentration of an electrolyte solution  are $\overline{d} = 3$, $V=0.3nm^3$, $n_{b}=0.1M$,  respectively.}
\label{fig:2}
\end{figure}

Fig. \ref{fig:2}(a) shows the spatial variation of the dimensionless electric potential near a charged soft surface, in the case where dimensionless  thickness of ion-penetrable polyelectrolytes and ionic concentration of an electrolyte solution are $\overline{d}=3$ and $n_{b}=0.1M$, respectively. Solid line and Plus signs stand for the cases having   $K_{\lambda} =1$ and $K_{\lambda}=0.5$ within our approach, respectively. Squares and Circles correspond to the cases having  $K_{\lambda} =1$ and $K_{\lambda}=0.5$ obtained by the approach of \cite%
{Das_PRE_2014} which does not consider orientational ordering of water dipoles.  Fig. \ref{fig:2}(a) shows that for the cases when considering orientational ordering of water dipoles,  the Donnan potentials are larger than when the orientational ordering of water dipoles is not considered. This can be proved by using the Eq. (\ref{eq:29}). 
\begin{eqnarray} 
	\frac{2e_{0}z n_{b}\sinh\left(\overline{\psi}_{D}\right)}{\left(1-2n_{b}V\right)\exp\left(\left(1-\frac{1}{\overline{V}}\right)\overline{h}\right)+2n_{b}V\cosh\left(\overline{\psi}_{D}\right)}=\frac{2e_{0}z n_{b}\sinh\left(\overline{\psi'}_{D}\right)}{\left(1-2n_{b}V\right)+2n_{b}V\cosh\left(\overline{\psi'}_{D}\right)},
\label{eq:30}
\end{eqnarray}
where $\overline{\psi'}_D$ is the Donnan potential for the case when the orientational ordering of water dipoles is not considered.
We rearrange Eq. (\ref{eq:30}) as follows 
\begin{eqnarray}
	\frac{2 n_{b}V}{1-2n_{b} V}\sinh\left(\overline{\psi}_{D}-\overline{\psi'}_{D}\right)=\exp\left(\left(1-\frac{1}{\overline{V}}\right)\overline{h}\right)\sinh\left(\overline{\psi'}_{D}\right)-\sinh\left(\overline{\psi}_{D}\right).
\label{eq:31}
\end{eqnarray}
If $\overline{\psi}_{D}<\overline{\psi'}_{D}$ then the left term is always less than 0, but the right term is unconditionally larger than 0.
In fact, we confirm that $\exp\left(\left(1-\frac{1}{\overline{V}}\right)\overline{h}\right)$ is always larger than 1 due to $\overline{h}>0$,$\overline{V}>1$.
In the case of  $\overline{\psi}_{D}=\overline{\psi'}_{D}$, then the left term is zero,but the right term is still larger than 0. For the only case of $V=V_{w}$, this condition $\overline{\psi}_{D}=\overline{\psi'}_{D}$ is satisfied .
However, for all the cases of $V\neq V_w$, it is proved that $\overline{\psi}_{D}>\overline{\psi'}_{D}$.
The difference in the electric potential is very small for the case of $K_{\lambda}=1$, but not for the case of $K_{\lambda}=0.5$. 
In physical point of view, the electrostatic potential should increase due to lowering of the permittivity attributed to orientational ordering of water dipoles. A large $K_{\lambda}$ means a small surface-charge-density, and then orientational odering of water dipoles gets very weak and hardly affects the screening property of the electrolyte solution.  

\begin{figure}
\includegraphics[width=1\textwidth]{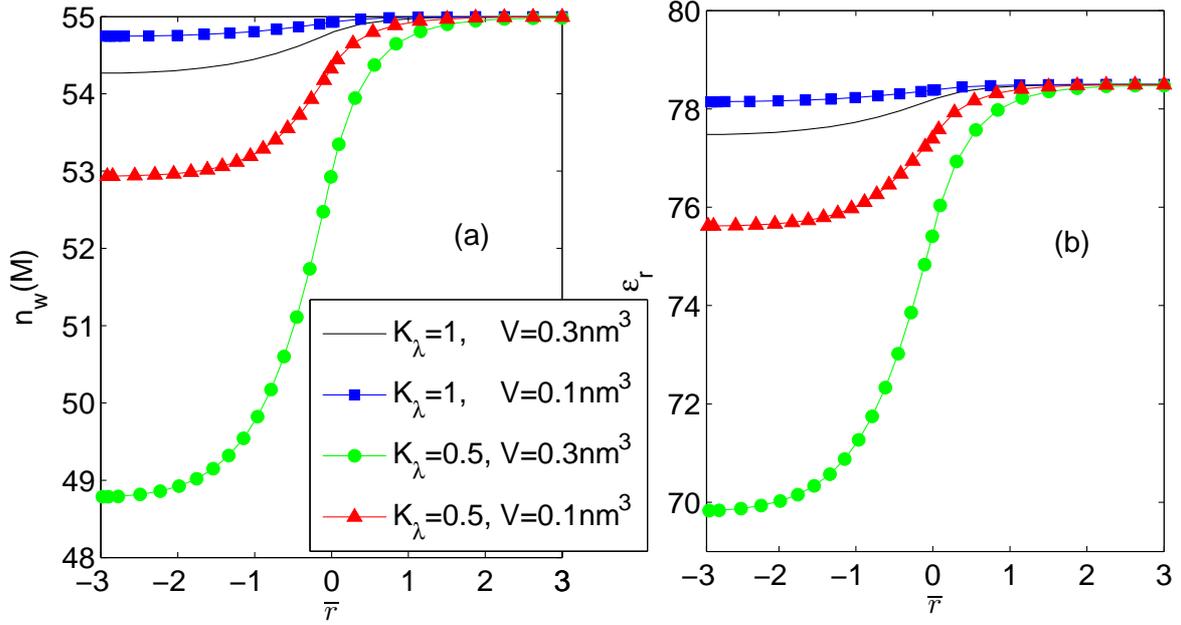}
\caption{\large (Color online) The number density of water molecules (a) and the permittivity of electrolyte solution (b) at $\overline{r}=-\overline{d}$  as a function of the distance from the soft charge surface. Other parameters are the same as in Fig. \ref{fig:2}.}
\label{fig:3}
\end{figure}
Fig. \ref{fig:2}(b) shows the variation of the dimensionless electric field strength with the position with respect to the interface between the FCL and the bulk electrolyte. As one can see, it is shown that the orientational ordering of water dipoles allows the electric field strength to enhance near the charged soft surface. This fact can be understood by the fact that accounting for orientational ordering of water dipoles lowers screening property of electrolytes and increases electric field strength everywhere near the charged soft surface.

Fig. \ref{fig:2}(c) shows the number density of coions as a function of the position with respect to the interface between the FCL and the bulk electrolyte.
Fig. \ref{fig:2}(c) shows that the number density of coions decreases with increasing the distance from the hard plate. It should be noted that when considering orientational ordering of water dipoles, the density is slightly smaller than when the orientational ordering of water dipoles is not considered. This is explained by combining  the effects of the term $\exp\left(\left(1-\frac{1}{\overline{V}}\right)\overline{h}\right)$ and the increase in the Donnan potential. Let's prove the fact.
The number densities of coions at the hard plate can be rewritten for different cases.
\begin{eqnarray}
	n_{co}=\frac{n_{b}\exp\left(-\overline{\psi}_{D}\right)}{\left(1-2n_{b}V\right)\exp\left(\left(1-\frac{1}{\overline{V}}\right)\overline{h}\right)+2n_{b}V\cosh\left(\overline{\psi}_{D}\right)}, n'_{co}=\frac{n_{b}\exp\left(-\overline{\psi'}_{D}\right)}{\left(1-2n_{b}V\right)+2n_{b}V\cosh\left(\overline{\psi'}_{D}\right)},
\label{eq:32}
\end{eqnarray}
where $n_{co}$ and $n'_{co}$ means the densities of coions at the hard plate for the cases when considering or not orientational ordering of water dipoles, respectively.
From the result of Fig. \ref{fig:2}(a), we know that  $\overline{\psi}_{D}>\overline{\psi'}_{D}$ and thus $\exp\left(-\overline{\psi}_{D}\right)<\exp\left(-\overline{\psi'}_{D}\right)$, $\cosh\left(\overline{\psi}_{D}\right)>\cosh\left(\overline{\psi'}_{D}\right)$.
On the other hand, the term $\exp\left(\left(1-\frac{1}{\overline{V}}\right)\overline{h}\right)$ is not less than 1.
As a result, $n_{co}\left(\overline{r}=-\overline{d}\right)<n'_{co}\left(\overline{r}=-\overline{d}\right)$.

\begin{figure}
\includegraphics[width=1\textwidth]{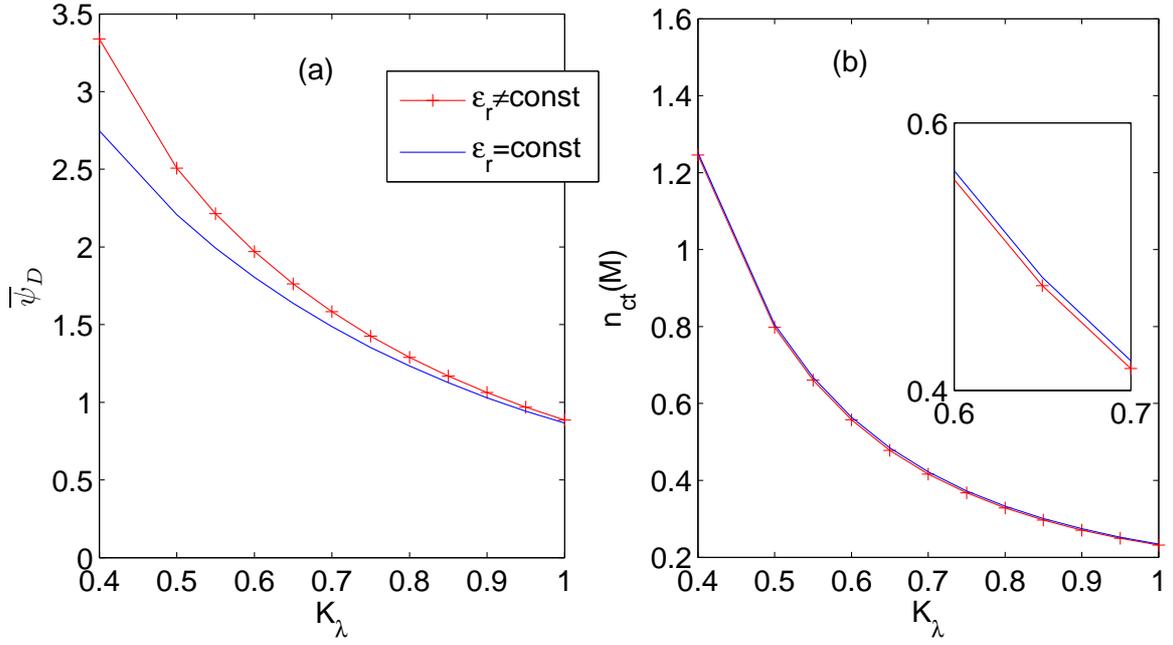}
\caption{\large (Color online) The dimensionless Donnan potential (a) and the number density of counterions at $\overline{r}=-\overline{d}$ (b) as a function of $K_{\lambda}$. The dimensionless thickness of ion-penetrable layer and the ion size are 3 and $0.3nm^3$, respectively. Other parameters are the same as in Fig. \ref{fig:2}. }
\label{fig:4}
\end{figure}
Fig. \ref{fig:2}(d) shows the number density of counterions  versus the distance from the charged soft surface. 
To neutralize the charge of FCL, the difference in the number densities of counterions and coions should be the same value when considering or not orientational ordering of water dipoles. Therefore, for the case when considering orientational ordering of water dipoles, the number density of counterions gets smaller than when the orientational ordering of water dipoles is not considered.

Fig. \ref{fig:3}(a) shows the number density of water dipoles as a function of the position with respect to the soft charge surface. Fig. \ref{fig:3}(a) shows that although for the case of low surface-charge-density($K_{\lambda}=1$), the number density of water molecules very slowly decreases with the distance from the soft charge surface, for the case of high surface-charge-density($K_{\lambda}=0.5$) the number densities of water molecules considerably decrease. It is also shown that a large ionic size ($V=0.3nm^3$) yields a smaller number density of water than for a relatively small size. This is a direct result from the fact that for large ions, screening the same quantity of soft charge requires more excluded volume of ions compared to original ones.  

Fig. \ref{fig:3}(b) shows the permittivity of an electrolyte solution versus the position with respect to the soft charge surface. Fig. \ref{fig:3}(b) represents that the permittivity decrease with decreasing of the distance from the hard plate. This fact is deduced from the decrease in number density of water molecules, shown in Fig. \ref{fig:3}(a). It can be straightforwardly understood from the expression of permittivity  Eq. (\ref{eq:20}).     

\begin{figure}
\includegraphics[width=1\textwidth]{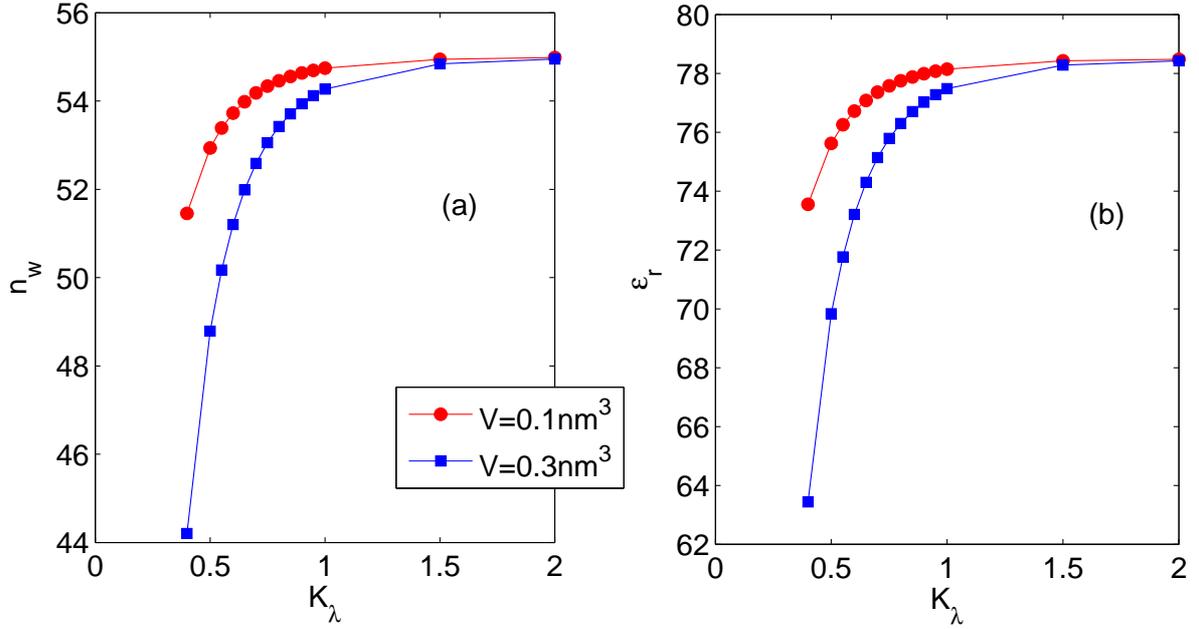}
\caption{\large (Color online) The number density of water molecules (a) and the permittivity of an electrolyte solution (b) at $\overline{r}=-\overline{d}$   as a function of $K_{\lambda}$.  Other parameters are the same as in Fig. \ref{fig:2}.}
\label{fig:5}
\end{figure}

Fig. \ref{fig:4}(a) shows the variation of the dimensionless Donnan potential with $K_{\lambda}$. It is shown that the increase in $K_{\lambda}$ results in the decrease of the Donnan potential. As shown in Fig. \ref{fig:2}(a), we again confirm that when considering the orientational ordering of water dipoles, for all $K_{\lambda}$, the Donnan potential is larger than when the orientational ordering of water dipoles is not considered. As $K_{\lambda}$ increases, the Donnan potentials  are close to each other because the effect of orientational ordering of water dipoles can be more and more negligible. 

Fig. \ref{fig:4}(b) shows the number density of counterions  at $\overline{r}=-\overline{d}$ as a function of $K_{\lambda}$ . 
As pointed out in Fig. \ref{fig:2}(d), for the case when considering orientational ordering of water dipoles, the number density of counterions is slightly smaller than when the orientational ordering of water dipoles is not considered.  It should be noted that the difference in the number densities is nearly constant.

Fig. \ref{fig:5}(a) shows the number density of water dipoles at the hard plate $\left(\overline{r}=-\overline{d}\right)$ of charged soft surface versus $K_{\lambda}$ for two values of $V=0.1, 0.3nm^3$. In the region of low $K_{\lambda}<1$, the number density of water dipoles rapidly increases with $K_{\lambda}$. This is attributed to the fact that an increase in $K_{\lambda}$ means a decrease in fixed charge and therefore within the FCL there exist a small number of electrolyte ions. An increase in $K_{\lambda}$ causes the number density of water molecules to increase towards the saturation value of $55mol/l$.  Due to the finite ion size effect, the number density of water molecules is lower for the large value of ionic size than for the relatively small one. 

\begin{figure}
\includegraphics[width=1\textwidth]{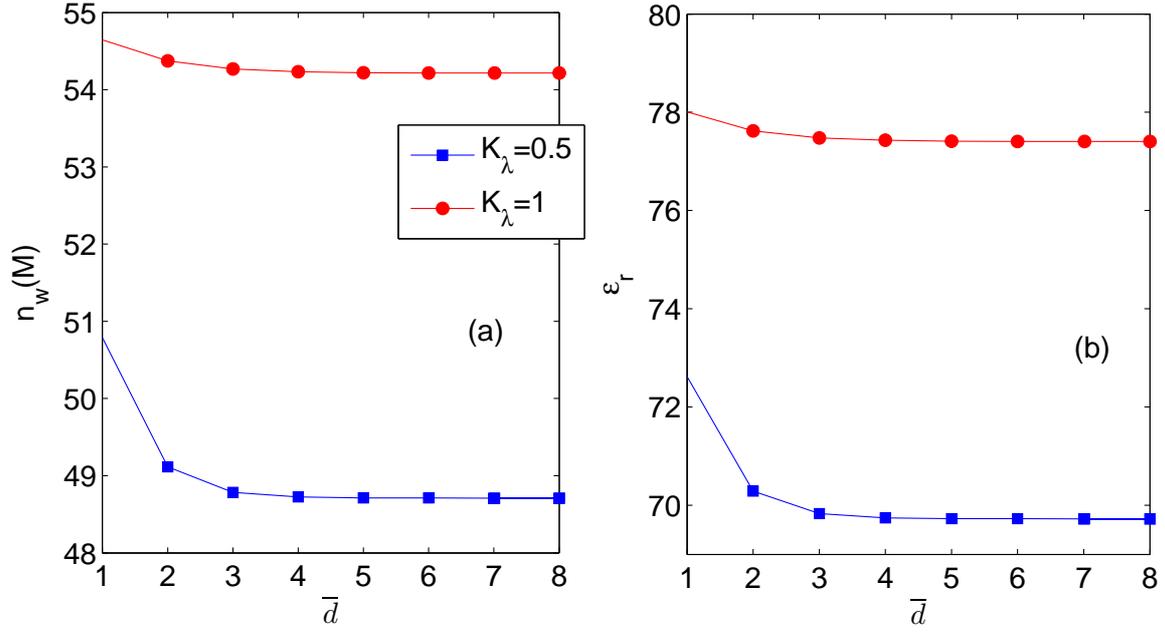}
\caption{\large (Color online) The number density of water molecules (a) and the permittivity of an electrolyte solution (b) at $\overline{r}=-\overline{d}$  as a function of the dimensionless thickness of ion-penetrable polyelectrolytes.  Other parameters are the same as in Fig. \ref{fig:2}. }
\label{fig:6}
\end{figure}
Fig. \ref{fig:5}(b) shows the permittivity of electrolyte solution at the hard plate  $\left(\overline{r}=-\overline{d}\right)$ of charged soft surface versus $K_{\lambda}$. 
Since the electric field strength is zero at the hard plate, the expression of the permittivity at the position can be simplified.
 The permittivity at the hard plate is proportional to the number density of water molecules according to the Eq. (\ref{eq:31}).  
\begin{eqnarray}
	\varepsilon_{r}\left(r=-d\right)=\lim_{r\rightarrow {-d}}\left(n^2+\frac{2+n^2}{3}p_{0}n_{w}\left(r\right)\frac{\mathcal{L}\left(\gamma p_{0}\beta E\right)}{\varepsilon_{0}E}\right)=n^2+\frac{2+n^2}{9}\frac{\gamma p_{0}^{2}\beta }{\varepsilon_{0}}n_{w}\left(r=-d\right)
\label{eq:31}
\end{eqnarray}

   Fig. \ref{fig:6}(a) shows the number density of water molecules at $\overline{r}=-\overline{d}$  as a function of the thickness d of ion-penetrable polyelectrolytes for two values of $K_{\lambda}=0.5, 1$. It is shown that for both cases, in the region of $\overline{d}<3$ the number density of water molecules decreases with $\overline{d}$. On the other hand, the number density of water molecules has a nearly same value everywhere in the region of $\overline{d}>3$. This can be understood from the fact that the increase in $\overline{d}$ results in constant number density of counterions near the hard plate\cite%
{Das_PRE_2014}. This also indicates that for the case when considering orientational ordering of water dipoles, the case of $\overline{d}=3$ also satisfies $\overline{d}>>\overline{\lambda}$, which is the necessary condition for Eq. (\ref{eq:29}). 

  Fig. \ref{fig:6}(b) shows number density of water molecules at $\overline{r}=-\overline{d}$  as a function of the thickness $\overline{d}$ of ion-penetrable polyelectrolytes. This behavior shows a linear relationship between the permittivity and the number density of water molecules at hard plate of charged soft surface.
Consequently, Fig. \ref{fig:6}(a) and Fig. \ref{fig:6}(b) mean that although for  $K_{\lambda} >1$, the effect of orientational ordering of water dipoles is negligible, the effect should be certainly considered  for $K_{\lambda} <0.5$. 
The present model is based on the assumption of an identical steric factor for both inside and outside the fixed charge layer, but the assumption is valid only for low grafting density of the fixed charge layer.  
The model can be amended by choosing non-uniform steric factor for the case when the grafting density of the fixed charge layer is medium or high. This improvement would change the Donnan potentials and the number densities of counterions and coions to more realistic values, respectively.

\section{Conclusions}
	In this work, we have presented a mean-field model accounting for the unequal sizes of ions and water molecules and orientational ordering of water dipoles in the electric double layer near a charged soft surface. We have derived the coupled equations for determining the number densities for ions and water dipoles in the electric double layer.  The consideration of orientational ordering of water dipoles allows the Donnan potential to increase compared to the case when orientational ordering of water dipoles is not considered.  On the other hand, the number densities of ions within FCL are slightly lower than when orientational ordering of water dipoles is not considered. The effect of orientational ordering of water dipoles is further enhanced if the ionic size increases.  In the near future, we will use this approach to describe the electrokinetics in soft nanochannel \cite%
{Das_JCIS_2015}, bacterial adhesion to surfaces \cite%
{Das_CSB_2015}, charging and swelling of cellulose films \cite%
{Das_JPCB_2015, Das_JPCB_2016},  the alteration in the conductivity of a biomembrane \cite%
{Casuso_Nanotech_2007, Webb_Biophys_1997}, regulation of osmotic pressure across a membrane \cite%
{Cath_EST_2009}, etc.

\end{document}